\documentclass[11pt,twoside]{article}
\usepackage{newpasp}

\usepackage{epsf}

\index{Pisano, D.J.}
\index{Wilcots, E.M.}

\begin{document}

\title{Assessing the state of galaxy formation.}
\author{D.J. Pisano}
\affil{Astronomy Dept., UW-Madison, 475 N. Charter St., Madison, WI 53706}
\author{Eric M. Wilcots}
\affil{Astronomy Dept., UW-Madison, 475 N. Charter St., Madison, WI 53706}

\begin{abstract}
We present the results of a survey of nearby, quiescent, non-peculiar, 
extremely isolated galaxies to search for the gaseous remnants of 
galaxy formation.  Such remnants are predicted to persist around galaxies
into the present day by galaxy formation models.  We find low-mass \HI\ 
companions around 7 of 34 galaxies surveyed.  In addition we find
5 galaxies with lopsided \HI\ distributions.  The implications for 
galaxy formation and the nature of high velocity clouds are discussed.
\end{abstract}



\section{Introduction}
What is the current state of galaxy formation in the local universe?  
There have been many recent detections of \HI\ clouds near larger spiral
galaxies in the local universe.  Such detections include \HI\ clouds 
around NGC 925 (Pisano, Wilcots, \& Elmegreen 1998), IC 10 (Wilcots \& Miller 
1998), four of five barred Magellanic spirals (Wilcots, Lehman, \& Miller 
1996), four of 16 low surface brightness dwarf galaxies and four of nine \HII\ 
galaxies (Taylor {\it{et al.}} 1993, 1996), and high-velocity clouds (HVCs) 
around M101 (Kamphuis 1993), NGC 628 (Kamphuis \& Briggs 1992, and in our own 
Local Group (see Wakker \& van Woerden 1997; Blitz {\it{et al.}} 1999).  
Typical clouds have 10$^7$-10$^8$\Msun\ of \HI\ amounting to 1\%-50\% of the 
mass of the primary galaxy.  

In numerous cases these \HI\ clouds have been suggested to be remnant material
from the galaxy formation process (e.g. NGC 925, IC 10, NGC 628, etc...).  
Current models of cold dark matter galaxy formation in which disk galaxies
were built up via the accretion of smaller bodies in a hierarchical merging 
process predict such remnant material to exist (e.g. Navarro, Frenk, \& 
White 1995).  Unfortunately the serendipitous nature of 
these detections inhibit our ability to divine the true origin of the \HI\
clouds.  These \HI\ clouds could be primordial material, but could also be
material ejected via a galactic fountain or superwind, tidal debris from
a recent interaction, or simply a small dwarf galaxy companion.  Therefore,
few, if any, of the \HI\ clouds represent unambiguous detections of the 
remnant reservoir of gas from which galaxies formed.  

To determine what the current state of galaxy formation is, we have 
conducted a systematic search for the remnant gas around a sample of 
extremely isolated and quiescent galaxies.  The results from the pilot
survey of six galaxies were reported in Pisano \& Wilcots 1999, here we
report of the current status of the expanded survey.

\section{Sample}
In order to determine the origin of \HI\ clouds around other galaxies it is
important to have a well-defined sample.  We chose galaxies from the Nearby
Galaxies Catalog (Tully 1988).  The galaxies were classified as 
isolated such that they had no known companions with $M_B\le$-16 mag within
1 Mpc of them.  In addition, galaxies were chosen that were classified as
non-peculiar.  These two conditions minimize the chance of the galaxy having
had a recent interaction or merger so there should be no tidal debris around 
our sample galaxies.  Our galaxies were also chosen to be quiescent (i.e.
not Seyferts or starbursts) so that any gas around these galaxies is unlikely
to be galactic ejecta from a galactic fountain or superwind.   

Finally, galaxies were chosen such that they were large enough and close
enough to resolve with the VLA in D configuration and the ATCA .750 
configuration (D$_{25}\ge$1\arcmin, R$\le$45 Mpc), yet far enough away to
probe out at least 90 kpc (R$\ge$21 Mpc).  This left us with 60 galaxies in
the entire sky; we observed 34 of those galaxies

\section{Observations}
Between November 1997 and February 2000 we observed a total of 34 galaxies
with the VLA and ATCA.  A total of 600 km s$^{-1}$ was covered at a resolution
of 5.2 km s$^{-1}$ for each observation.  The sample galaxies had distances
between 21 Mpc and 45 Mpc, allowing us to survey out to a radius of 92 - 192 
kpc at a resolution of $\sim$1\arcmin  (6.1-12.8 kpc).  The resulting 
observations have $\sigma\simeq$0.5-1$\times$10$^{19}$cm$^{-1}$ per channel
for the column density.  The mass detection limits are 
8.3$\pm$3.5$\times$10$^6$\Msun\ for a 5$\sigma$ detection over 2 channels (10.4
km s$^{-1}$).  The range of mass detection limits comes from varying 
sensitivity and distance for each galaxy.

\section{Results}
Of the 34 galaxies surveyed we detected gas-rich companions in \HI\ around 
7 of them (see figures 1 \& 2).  Another 5 galaxies have ``disturbed''
\HI\ morphologies (figures 3 \& 4); either severe warps or lopsided 
distributions possibly indicative of a recent minor merger.  
The remaining 22 galaxies are relatively normal with all of 
the idiosyncrasies we typically see in galaxies such as small warps and 
asymmetries.  

\begin{figure}
\caption{UGC 11152, an isolated galaxy accreting a gas-rich companion.  The
left panel is total \HI\ intensity on an optical image from the Digital Sky
Survey with contours starting at 10$^{19}$cm$^{-2}$ with increments of a half a
dex.  The right panel is the \HI\ velocity field on top of the \HI\ total 
intensity.  Some velocity contours are labeled for reference.  Contours are 
spaced by 20 km s$^{-1}$.}
\end{figure}

\begin{figure}
\caption{NGC 2708, an isolated galaxy interacting with a gas-rich companion.  
Panels are as in figure 1.}
\end{figure}

\begin{figure}
\caption{IC 5078, an isolated galaxy with a ``disturbed'' \HI\ morphology.  
Panels are as in figure 1.}
\end{figure}

\begin{figure}
\caption{NGC 895, a severely warped isolated galaxy.  Panels are as in figure 1.}
\end{figure}

The detected companions have M$_{\HI}$ between 10$^8$\Msun\ and 10$^9$\Msun,
which corresponds to 3\%-30\% of the primary galaxy's mass in \HI.  The 
companions appear show signatures of rotation, so based
on the rotation widths and sizes of the companions we determined their
dynamical masses to be between 10$^9$\Msun\ and 10$^{10}$\Msun, which is
0.5\%-10\% of the main galaxy's dynamical masses.  The ratio of \HI\ mass to
dynamical mass for the companions range from 7\%-85\%.  All \HI\ clouds
detected have spatially coincident optical emission, with the possible 
exception of UGC 11152.  All of these properties are consistent with the
gas-rich companions being typical dwarf galaxies.  

This does not, however, mean that we have not detected the gaseous remnants
of galaxy formation, but simply that these gas clouds formed stars before 
being accreted by the primary galaxy.  It is important to confirm, however, 
that these gas-rich companions will eventually be accreted.  Our companions
have projected separations of 20-100 kpc (1-6 R$_{gal}$) in radius and 20-100
km s$^{-1}$ in velocity.  These numbers imply an orbit time, which is 
roughly equal to the dynamical friction timescale, of 5-10 Gyr.  These
companions turn out to be in relatively stable orbits.  

\section{Implications for the nature of High-Velocity Clouds}

The Blitz {\it{et al.}} (1999) model for the origin of HVCs in the Local
Group suggests that they are primordial material left over from the 
formation of the Local Group.  In this model HVCs are at large distances
from the Milky Way (R$\sim$750kpc-1Mpc) and, therefore, have large masses
(M$_{\HI}\sim$10$^7$\Msun).  

While our survey was not optimized to examine the origin of HVCs, there are
some intriguing implications from its results.  We detected no objects that
resembled HVCs (i.e. no gas-rich companions without stars).  Furthermore, 
we detected no companions smaller than 10$^8$\Msun\ in \HI\ down to our 
detection limit at $\sim$10$^7$\Msun.  This implies that if HVCs are 
associated
with galaxy formation, they must either have masses lower than 10$^7$\Msun\
and/or be at projected separations greater than 140 kpc.  The former suggestion
is somewhat unlikely, because we do detect larger companions and while one
might expect more \HI\ clouds at lower masses, we do not detect any down to 
our detection limit.  Another possible explanation for our non-detection of 
HVCs is that they are associated with group formation, and not with the 
formation of individual, isolated galaxies.  Either way, any explanation for 
a primordial origin of HVCs in the Local Group must account for our 
non-detection of them around isolated galaxies.

\section{Implications for Galaxy Formation}

The main goal of this work was to assess the state of galaxy formation in
the local universe.  At this point in time, this work has yielded three main
implications for galaxy formation:

First, galaxy formation appears to be an efficient process.  As discussed 
above, we found no companions with \HI\ masses below 10$^8$\Msun\ implying 
that there is not a population of low mass \HI\ clouds within $\sim$100 kpc of 
these galaxies.  

Second, galaxy formation has basically concluded.  Only $\sim$20\% of isolated
galaxies have low-mass ($\sim$10\%M$_{gal}$), gas-rich companions, (which
are in stable orbits).  Therefore most of the gas must already be in the 
main galaxy.  The mass outside of the main galaxy will take a long time
($\sim$5 Gyr) to be accreted.

Third, galaxy formation may have recently ended.  Another 15\% of isolated
galaxies have disturbed \HI\ morphologies suggesting that they may have 
recently (in the last 1 Gyr) undergone a minor merger ($\le$10\% M$_{gal}$).

The accretion rates implied by this work (10\% of M$_{gal}$ over 6 Gyr for 
$\sim$40\% of galaxies) are consistent with those derived by Toth \& Ostriker
(1992) from the scale height of the Milky Way, Zaritsky \& Rix (1997) from 
the frequency of asymmetries, and Navarro, Frenk, \& White (1995) from 
galaxy formation simulations.  Future work on this project will involve
further increasing the number of galaxies surveyed, comparing the stellar
properties of the main galaxies and companions with their \HI\ properties and 
more detailed comparisons of our detection rates with theoretical predictions.

\acknowledgments
D.J.P. wishes to thank the Wisconsin Space Grant Consortium for support for 
this work and the conference organizers for providing a student travel grant
to help attend the conference.  D.J.P. and E.M.W. were supported by NSF grants
AST-9616907 and AST-9875008.

\end{document}